\documentclass[english]{acmart}
\makeatletter
\def\@ACM@checkaffil{% Only warnings
    \if@ACM@instpresent\else
    \ClassWarningNoLine{\@classname}{No institution present for an affiliation}%
    \fi
    \if@ACM@citypresent\else
    \ClassWarningNoLine{\@classname}{No city present for an affiliation}%
    \fi
    \if@ACM@countrypresent\else
        \ClassWarningNoLine{\@classname}{No country present for an affiliation}%
    \fi
}
\makeatother
\usepackage[utf8]{inputenc}
\usepackage{graphicx}

\makeatletter
\renewcommand\footnotetextcopyrightpermission[1]{} % removes footnote with conference information in first column
\setcopyright{none}
\makeatletter
\renewcommand\@formatdoi[1]{\ignorespaces}
\makeatother

\usepackage[frozencache,cachedir=_minted-nelli]{minted}
\definecolor{bg}{rgb}{0.95,0.95,0.95}
\setminted{
    bgcolor=bg,
    escapeinside={||},
    mathescape=true,
	frame=single,
	autogobble=true,
}
\setmintedinline{breaklines,escapeinside={||},mathescape=true}
\AtBeginEnvironment{minted}{%
  \catcode`?\active
  \begingroup\lccode`~=`\?\lowercase{\endgroup\def~{\linebreak}}%
}

\usepackage{float} %minted inside figure
\usepackage{url}

\usepackage{tikz}
\usetikzlibrary{calc,tikzmark}
\colorlet{colorMain}{green} 
\colorlet{colorSum}{yellow}

\makeatother

\usepackage{babel}
\usepackage{minted}

\begin{document}
\author{Maksim Levental}
\email{mlevental@uchicago.edu}
\affiliation{%
  \institution{University of Chicago}
  \country{USA}
}
\author{Alok Kamatar}
\email{alokvk2@uchicago.edu}
\affiliation{%
  \institution{University of Chicago}
  \country{USA}
}

\author{Ryan Chard}
\email{rchard@anl.gov}
\affiliation{%
  \institution{Argonne National Laboratory}
  \country{USA}
}
% \author{Nicolas Vasilache}
% \email{ntv@google.com}
% \affiliation{%
%   \institution{Google Brain}
%   \country{Switzerland}
% }
% \author{Ingo Müller}
% \email{ingomueller@google.com}
% \affiliation{%
%   \institution{Google Brain}
%   \country{Switzerland}
% }
\author{Kyle Chard}
\email{chard@uchicago.edu}
\affiliation{%
  \institution{University of Chicago}
  \country{USA}
}
\author{Ian Foster}
\email{foster@uchicago.edu}
\affiliation{%
  \institution{University of Chicago}
  \country{USA}
}

\title{\texttt{nelli}: A lightweight frontend for MLIR}

\begin{abstract}
Multi-Level Intermediate Representation (MLIR) is a novel compiler infrastructure that aims to provide modular and extensible components to facilitate building domain specific compilers. 
MLIR enables modeling programs across the spectrum of abstraction levels.
However, since all existing frontends are at a very high level of abstraction, the semantics and mechanics of the fundamental transformations available in MLIR are difficult to investigate and employ per se.
To address these challenges, we have developed \texttt{nelli}, a lightweight, Python-embedded, domain-specific, language for generating MLIR representations. 
\texttt{nelli} leverages existing MLIR infrastructure to develop Pythonic syntax and semantics for various MLIR features. 
We describe \texttt{nelli}'s design goals, discuss key details of our implementation, and demonstrate how \texttt{nelli} enables easily defining and lowering compute kernels to diverse hardware platforms.
\end{abstract}
\maketitle
\tableofcontents{}

\section{Introduction\label{sec:Introduction}}

MLIR is a modular and extensible compiler infrastructure \citep{https://doi.org/10.48550/arxiv.2002.11054} for progressively transforming (\emph{lowering}) programs from high-level (in terms of abstraction), architecture-independent representations to low-level, architecture-specific representations. 
Such Intermediate Representations (IRs) are termed
\emph{dialects} in the MLIR context in order to emphasize their mutual compatibility and the unified interface that MLIR provides for transforming between them, a process referred to as \emph{running passes}.

MLIR has been applied to various problem domains and in its default distribution (other, so-called ``out-of-tree,'' implementations exist) supports representing programs ranging in abstraction level from dataflow compute graphs, such as can be used to represent Deep Neural Networks (DNNs), to architecture-specific vector instructions. 
Other less quotidian applications of MLIR include modeling gate-level primitives \citep{eldridge2021mlir}, quantum assembly languages \citep{9605269}, and database query languages
\citep{blockhaus2022framework,10.14778/3551793.3551801}. 
By virtue of its close connection to LLVM \citep{lattner2004llvm},
MLIR supports code generation for CPUs, GPUs, and other compute platforms, including abstract runtimes (as opposed to concrete hardware architectures) such as OpenMP.

While the primary value of MLIR is its support for efficient (quick) construction of IRs modeling novel domains, an undeniable secondary value is the ability to use existing dialects, corresponding to established programming
models, in combination with novel transformations tailored to problem-specific hardware configurations. 
For example, while there was been much research on the use of MLIR to lower DNNs to high-performance CPU and GPU platforms \citep{https://doi.org/10.48550/arxiv.2202.03293},
such as data-center class devices and high-powered mobile devices (e.g.,
expensive mobile phones), there is a dearth of work on efficiently
targeting low-power edge devices, such as micro-controllers and single-board computers. 
Yet those latter edge devices, while relatively underpowered, can be an attractive DNN deployment target in instances where power is a scarce commodity, such as IoT,
AgTech, and urban infrastructure monitoring. 
Indeed, it is conceivable that,
given sufficient directed design space exploration (such as can be
realized by using MLIR), these low-power edge devices could effectively
support edge inference of DNNs.

However, while MLIR provides the edge device software architect with a rich existing repository of useful dialects and transformations, the effective use of those capabilities for edge device programming is hindered by the lack of a point of ingress to MLIR capabilities that is not encumbered by assumptions about the roles of the existing dialects and their mutual relationships.
Specifically, almost all extant ingress points take the form of high-level DNN frameworks, such as PyTorch \citep{paszke2017automatic}, TensorFlow
\citep{https://doi.org/10.48550/arxiv.1603.04467}, or ONNX \citep{https://doi.org/10.48550/arxiv.2008.08272}---but 
most optimization actually occurs on lower-level dialects, such as the affine, structured control-flow, and vector dialects. 
Thus, in order to productively investigate possible optimization
opportunities one must distinguish artifacts of the lowering process
from the kernel representations themselves. 
For example, consider investigating the optimization of a ($32\times32$) linear layer (i.e., \mintinline{python}{torch.nn.Linear(32, 32)}).
This ubiquitous DNN operation lowers to the loop nests in Listing
\ref{listing1}; note that the third loop nest is readily identified as
corresponding directly to a matrix-multiplication kernel, but the other
three are somewhat mysterious\footnote{Case in point: the seemingly redundant initialization and subsequent
copy into an intermediate buffer in the lowering of \mintinline{python}{torch.nn.Linear(32, 32)}
is the result of \mintinline{mlir}{torch-mlir} enforcing value semantics
\citep{smith2002introducing} on \mintinline{mlir}{torch.tensor}s,
which, while important, obscures the actual compute kernel.}. Thus, in longer programs (complete DNNs) it becomes difficult to
identify, isolate, and manipulate (e.g., to optimize) IR corresponding
most closely to the compute kernel itself, amongst IR that reflects
certain assumptions/contracts. 
Conversely, there currently exists
no simple and efficient way to emit any of the lower-level dialects
in MLIR (such as \mintinline{mlir}{scf}, \mintinline{mlir}{affine},
\mintinline{mlir}{memref}, or \mintinline{mlir}{vector}) short of
writing the IR ``by hand.''

\medskip{}

\begin{figure}
\centering{}\renewcommand\figurename{Listing.}%
\begin{minipage}[t]{0.45\textwidth}%
\begin{minted}[tabsize=4]{python}
@mlir_func
def ifs(M: F64, N: F64):
    one = 1.0
    if M < N:
        two = constant(2.0)
        mem = MemRef.alloca([3, 3], F64)
    else:
        six = constant(6.0)
        mem = MemRef.alloca([7, 7], F64)
	return
\end{minted}
\end{minipage}\hfill{}%
\begin{minipage}[t]{0.53\textwidth}%
\begin{minted}[tabsize=4]{mlir}
func.func @ifs(%M: f64, %N: f64) {
  %one = arith.constant 1.000000e+00 : f64
  %cond = arith.cmpf olt, %arg0, %arg1 : f64
  scf.if %cond {
    %two = arith.constant 2.000000e+00 : f64
    %mem = memref.alloca() : memref<3x3xf64>
  } else {
    %six = arith.constant 6.000000e+00 : f64
    %mem = memref.alloca() : memref<7x7xf64>
  }
  return
}
\end{minted}
\end{minipage}\caption{\texttt{nelli} mapping between Python's \protect\mintinline{python}{if}
and MLIR's \protect\mintinline{mlir}{scf} dialect.}
\label{nelli_if_to_scf_if}
\end{figure}

\begin{figure}
\centering{}\renewcommand\figurename{Listing.}%
\begin{minipage}[t]{0.35\textwidth}%
\begin{minted}[tabsize=4]{python}
M, N, K = 4, 16, 8

@mlir_func
def matmul(
	A: MemRef[(M, N), F32], 
	B: MemRef[(N, K), F32], 
	C: MemRef[(M, K), F32]
):
    for i in range(M):
        for j in range(N):
            for k in range(K):
                a = A[i, j]
                b = B[j, k]
                c = C[i, k]
                d = a * b
                e = c + d
                C[i, k] = e
\end{minted}
\end{minipage}\qquad{}%
\begin{minipage}[t]{0.45\textwidth}%
\begin{minted}[tabsize=4]{mlir}
func.func @matmul(
	%A: memref<4x16xf32>, 
	%B: memref<16x8xf32>, 
	%C: memref<4x8xf32>
) {
  affine.for %i = 0 to 4 {
    affine.for %j = 0 to 16 {
      affine.for %k = 0 to 8 {
        %a = memref.load %A[%i, %j] 
        %b = memref.load %B[%j, %k]
        %c = memref.load %C[%i, %k]
        %d = arith.mulf %a, %b : f32
        %e = arith.addf %c, %d : f32
        memref.store %e, %C[%i, %k] 
      }
    }
  }
  return
}
\end{minted}
\end{minipage}\caption{\texttt{nelli} mapping between Python's \protect\mintinline{python}{for}
and MLIR's \protect\mintinline{mlir}{affine} dialect.}
\label{nelli_for_to_affine_for}
\end{figure}

In order to address the problem of MLIR's lower-level dialects being inaccessible, we present \texttt{nelli}\footnote{\url{https://github.com/makslevental/nelli}},
a lightweight frontend for MLIR. 
This Python embedded
domain-specific language (eDSL) builds on top of existing MLIR
Python bindings to map Python primitives (such as \mintinline{python}{if}s,
\mintinline{python}{for}s, and \mintinline{python}{class}es) to
various MLIR dialects. 
Our foremost goal in designing \texttt{nelli} was to make
MLIR more ergonomic and thereby more accessible. 
To this end, we make \texttt{nelli} \textquotedbl Pythonic\textquotedbl{} while preserving MLIR semantics vis-a-vis the in-tree Python bindings. 
See Listings \ref{nelli_if_to_scf_if} and \ref{nelli_for_to_affine_for}
for some examples of \texttt{nelli} syntax. 
Notably, \texttt{nelli} captures program control flow and produces fully typed IR with little static analysis on the Python source (hence,
\textit{lightweight}).
Additionally, since \texttt{nelli} is a Python
eDSL, it fully interoperates with existing Python tooling (IDEs, debuggers,etc.) and other elements of the Python ecosystem. 

In the following, we discuss in greater detail \texttt{nelli} design goals, the eDSL implementation approaches that we investigated, and the implementation details of our chosen approach.
We also present three use cases: 1) a kernel tuner that uses a black-box,
gradient-free, optimizer \citep{nevergrad}, demonstrating the power and convenience of Python interoperability;
2) a pipeline for lowering kernels to target GPUs and then evaluating performance on a Raspberry Pi edge device, demonstrating ease
of integration with LLVM, downstream of MLIR; 3) and a pipeline for translating parallelizable kernels to OpenMP programs.
In summary, this work makes the following contributions:
\begin{enumerate}
\item A thorough discussion of several alternative eDSL implementation approaches
(in Python) and their relative merits and deficiencies.
\item A discussion of the design and implementation of an embedded domain-specific
language (\texttt{nelli}) with minimal static (ahead-of-time) analysis
and complexity; 
\item Implements several lowerings that demonstrate capabilities of \texttt{nelli},
with a focus on deploying compute intensive kernels to diverse hardware
platforms.
\end{enumerate}
The remainder of the paper is structured as follows: Section \ref{sec:Background}
reviews the relevant background on eDSLs and MLIR; Section \ref{sec:Implementation}
discusses the implementation of \texttt{nelli}; Section \ref{sec:Demonstration-and-Evaluation}
demonstrates the capabilities of \texttt{nelli}; and, finally, Section
\ref{sec:Related-Work} compares \texttt{nelli} to similar tools.

\section{Background\label{sec:Background}}

We quickly review the necessary background on MLIR, in particular
with respect to DNN deployment to edge devices, and eDSL construction
in general.

\subsection{MLIR\label{subsec:MLIR}}

\begin{figure}
\begin{centering}
\includegraphics[scale=0.75]{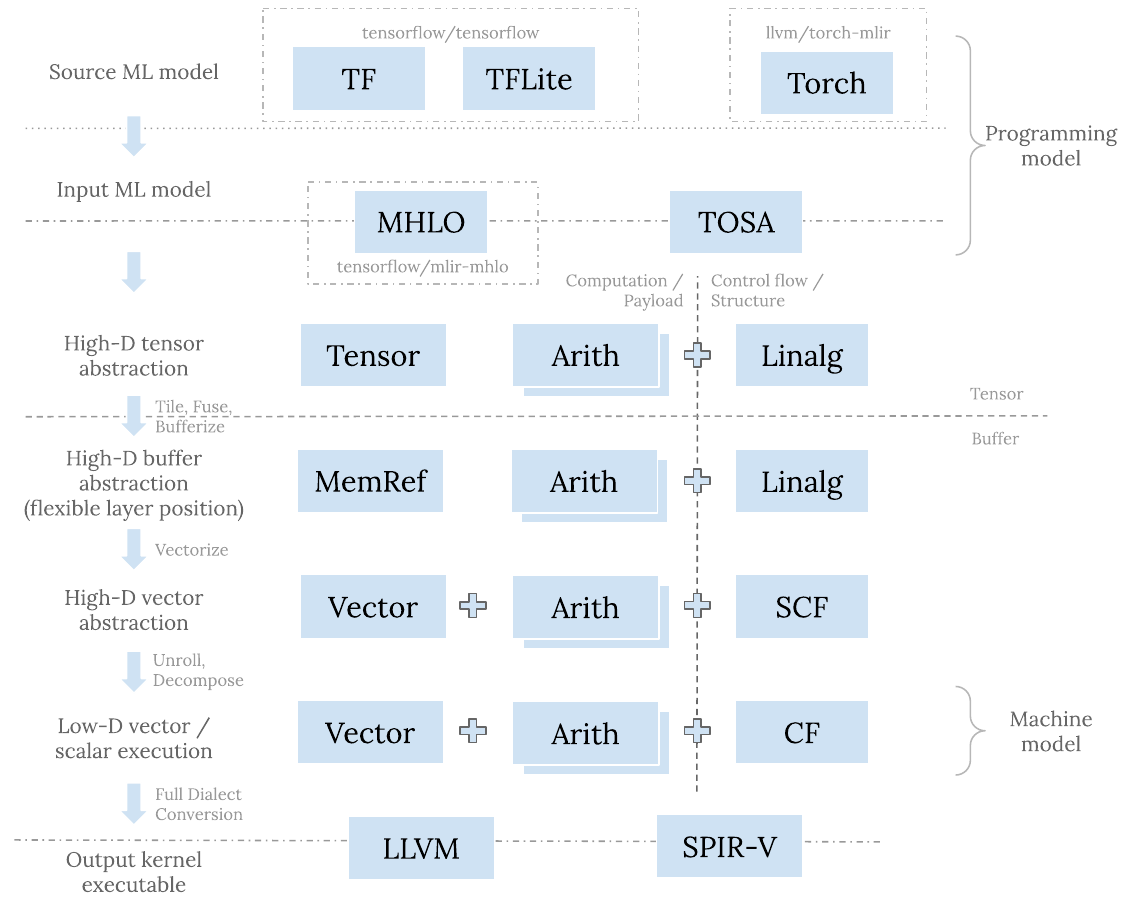}
\par\end{centering}
\caption{Ladder of dialect abstraction in terms of dialect types and dialect
operations (reprinted with permission from \citep{ladderdialect});
with respect to types, a progressive lowering needs representations
for tensors, buffers, vectors, and scalars, while with respect to
operations, it needs to support computation/payload (i.e., arithmetic)
and control flow.}
\label{ladderofdialectabstraction}

\end{figure}

MLIR is an approach to building reusable and extensible compiler infrastructure.
Practically this means that MLIR constitutes a collection of utilities
for 
\begin{enumerate}
\item Defining mutually compatible IRs, known as \emph{dialects}, that model
programs in particular domains, supporting operations (including attributes)
and types (including traits);
\begin{enumerate}
\item[$\bullet$] Using the Operation Definition Specification (ODS) language implemented
against LLVM's TableGen\footnote{MLIR is an ``in-tree'' LLVM project and thus reuses and extends
many of LLVM's existing facilities.} utility;
\end{enumerate}
\item Defining intra-dialect transformations, such as canonicalization,
inlining, and dead-code elimination;
\begin{enumerate}
\item[$\bullet$] Using a subgraph matching\footnote{Directed, acyclic, graph matching is strictly more powerful than tree
matching \citep{ebner2008generalized}.} and rewriting concept known as a \mintinline{cpp}{RewritePattern};
\end{enumerate}
\item Defining inter-dialect transformations, known as \emph{conversions;}
\begin{enumerate}
\item[$\bullet$] Using \mintinline{cpp}{ConversionPattern}s and \mintinline{cpp}{TypeConverter}s.
\end{enumerate}
\end{enumerate}
In addition to a thriving ecosystem of dialects, tools, and down-stream
projects, MLIR has many ``in-tree'' dialects that model programs
across the abstraction-level spectrum. It also supports target-specific
code generation and runtime execution through the various backends
provided by LLVM; this includes both \texttt{x86\_64} and \texttt{aarch64}/\texttt{arm64}
CPU instruction set architectures (ISAs), NVPTX\footnote{NVIDIA Parallel Thread Execution is a virtual machine instruction
set architecture used by NVIDIA's GPUs as an interface layer between
CUDA and SASS; SASS is the low-level assembly language that compiles
to binary microcode, which executes natively on NVIDIA GPU hardware
\citep{nvidiasass}.} and SPIR-V\footnote{Khronos Group\textquoteright s binary intermediate language SPIR-V
for representing graphics shaders and compute kernels \citep{kessenich2018spir};
both the Vulkan graphics API and the OpenCL compute API support SPIR-V.} GPU pseudo-ISAs, as well as minimal runtimes for each. See Figure
\ref{ladderofdialectabstraction} for the ``ladder of abstraction''
\citep{hayakawa1948art} in terms of MLIR dialects. We briefly describe
a few of the dialects (in-tree and out-of-tree) relevant for DNN deployment
to edge devices.

\subsubsection{High-level dialects}

At the highest level of abstraction, MLIR supports representing DNNs
specified using high-level frameworks such as TensorFlow and PyTorch\footnote{The \mintinline{mlir}{tf} (TensorFlow), \mintinline{mlir}{tfl} (TensorFlowLite),
\mintinline{mlir}{torch} (PyTorch), and \mintinline{mlir}{mhlo}
dialects are all ``out-of-tree'' dialects.}. The role of these dialects (\mintinline{mlir}{tf}, \mintinline{mlir}{tfl},
\mintinline{mlir}{torch}) is to faithfully represent the source model
as specified in the chosen framework and thus they function as points
of ingress into MLIR. As mentioned in the introduction, this effectively
makes TensorFlow and PyTorch the only mature points of ingress into
MLIR (see Section \ref{sec:Related-Work}). In addition,
as evinced by Listings \ref{torchnnconv2dpytorch} and \ref{torchnnconv2dtorchdialect},
the lowering/translation process incurs a high cost with respect to
legibility; naturally, lowering the level of abstraction necessitates
the inclusion of explicit specification of operations that are implicit
(or, at least, taken for granted) in the high-level representation.
With Listings \ref{torchnnconv2dpytorch} and \ref{torchnnconv2dtorchdialect}
in mind, note that in specifying the operation \mintinline{python}{torch.nn.Linear(32, 32)},
one implicitly specifies:
\begin{enumerate}
\item The bias tensor \mintinline{mlir}{|\%|0 = torch.vtensor.literal -> !torch.vtensor<|{[}|32|{]}|,f32>}
needs to be initialized;
\item The weight tensor \mintinline{mlir}{|\%|1 = torch.vtensor.literal -> !torch.vtensor<|{[}|32, 32|{]}|,f32>}
needs to be initialized; 
\item A transpose \mintinline{mlir}{|\%|2 = torch.aten.transpose.int |\%|1, |\%|int0, |\%|int1}
on the weight tensor needs to be performed (since, in PyTorch, weights
are stored in column-order);
\item The bias needs to be added \mintinline{mlir}{|\%|4 = torch.aten.add.Tensor |\%|3, |\%|0, |\%|float1.0e00}
to the result of the matrix multiplication (\mintinline{mlir}{|\%|3 = torch.aten.mm |\%|arg0, |\%|2}).
\end{enumerate}
\begin{figure}
\renewcommand\figurename{Listing.}%
\begin{minipage}[t]{0.37\textwidth}%
\begin{minted}[tabsize=4]{python}
class MyMatmul(nn.Module):
    def __init__(self):
        super().__init__()
		self.matmul = nn.Linear(
			32, 32
		)

    def forward(self, x):
        return self.matmul(x)
\end{minted}
\caption{Simple neural network with a \protect\mintinline{python}{torch.nn.Linear(32, 32)}.}
\label{torchnnconv2dpytorch}%
\end{minipage}\hfill{}%
\begin{minipage}[t]{0.6\textwidth}%
\begin{minted}[tabsize=4]{mlir}
module attributes {
  func.func @forward(
	%arg0: !torch.vtensor<[32,32],f32>
	) -> !torch.vtensor<[32,32],f32> {
    %int0 = torch.constant.int 0
    %int1 = torch.constant.int 1
    %float1.0e00 = torch.constant.float 1.0
    %0 = torch.vtensor.literal(dense<1.0>) 
    %1 = torch.vtensor.literal(dense<1.0>)
    %2 = torch.aten.transpose.int %1, %int0, %int1 
    %3 = torch.aten.mm %arg0, %2
    %4 = torch.aten.add.Tensor %3, %0, %float1.0e00
    return %4 : !torch.vtensor<[32,32],f32>
  }
}
\end{minted}
\caption{Simple neural network with a \protect\mintinline{python}{torch.nn.Linear(32, 32)}
layer rendered in the \protect\mintinline{mlir}{torch} dialect.}
\label{torchnnconv2dtorchdialect}%
\end{minipage}
\end{figure}

The ultimate effect of this translation process is that targeting
the operation of interest (e.g., \mintinline{python}{torch.nn.Linear(32, 32)})
for investigation and transformation is made more difficult. It's
important to not underestimate the significance of the last point:
subgraph matching, as implemented in MLIR by the \mintinline{cpp}{RewritePattern},
``anchors'' on a target operation. Thus, if an optimizing transformation
(such as loop-unrolling, loop-fusion, loop-tiling) is implemented
targeting the loop nests generated from this high-level representation
(see Listing \ref{listing1}), then running that pass will incur high(er)
runtime cost\footnote{Imagine targeting the third loop-nest in Listing \ref{listing1};
you might develop a \mintinline{cpp}{RewritePattern} that matches
on \mintinline{mlir}{scf.for} but then there are $2+2+3+2=9$ possible
such matches. Thus, one needs to further filter the possible matches
(e.g., by filtering on whether the body contains a sequence of \mintinline{mlir}{arith.mulf}
and \mintinline{mlir}{arith.addf}).} and much higher development time. In MLIR, the partial resolution
to this problem is called \emph{structured code generation} \citep{https://doi.org/10.48550/arxiv.2202.03293},
i.e., high-level operations such as \mintinline{python}{torch.nn.Linear(32, 32)}
are first lowered to a structured representation, such as \mintinline{mlir}{linalg.generic}
(see Listing \ref{linalg-matmul}), which is itself transformed and
lowered to optimized loop-nests (see Listing \ref{listing1}). But,
as can be observed in Listing \ref{linalg-matmul}, these structured
transformations are (currently) limited to kernels implemented in
terms of \texttt{parallel} and \texttt{reduction} iterators.

\begin{figure}
\centering{}%
\begin{minipage}[t]{0.65\textwidth}%
\begin{minted}[tabsize=4,highlightlines={14-26}]{mlir}
#map3 = affine_map<(d0, d1, d2) -> (d0, d2)>
#map4 = affine_map<(d0, d1, d2) -> (d2, d1)>
#map5 = affine_map<(d0, d1, d2) -> (d0, d1)>
%3 = linalg.generic {
  indexing_maps = [#map3, #map4, #map5],
  iterator_types = ["parallel", "parallel", "reduction"]
} ins(%arg0, %1 : tensor<32x32xf32>, tensor<32x32xf32>)
  outs(%2 : tensor<32x32xf32>) {
^bb0(%in: f32, %in_2: f32, %out: f32):
  %5 = arith.mulf %in, %in_2 : f32
  %6 = arith.addf %out, %5 : f32
  linalg.yield %6 : f32
} -> tensor<32x32xf32>
\end{minted}
\end{minipage}\caption{\protect\mintinline{mlir}{linalg.generic} representation for \protect\mintinline{python}{torch.nn.Linear(32, 32)}.}
\label{linalg-matmul}
\end{figure}

\subsubsection{Intermediate-level dialects}

An intermediate-level dialect
is one that can be used to represent a kernel explicitly but is abstract
with respect to hardware implementation. Thus, the structured control
flow dialect (\mintinline{mlir}{scf}), which models loops (\mintinline{mlir}{scf.for},
\mintinline{mlir}{scf.while}, \mintinline{mlir}{scf.parallel}),
and the \mintinline{mlir}{memref} dialect, which is intended to model
creation and manipulation of objects with memory reference semantics
(i.e., buffers). See Listing \ref{listing1} for the representation
of \mintinline{python}{torch.nn.Linear(32, 32)} purely in terms of
these dialects.

\begin{figure}
\centering{}%
\begin{minipage}[t]{0.75\textwidth}%
\begin{minted}[tabsize=4,linenos,highlightlines={14-26}]{mlir}
%alloc = memref.alloc() {alignment = 64 : i64} : memref<32x32xf32>
scf.for %arg1 = %c0 to %c32 step %c1 {
  scf.for %arg2 = %c0 to %c32 step %c1 {
    memref.store %cst, %alloc[%arg1, %arg2] : memref<32x32xf32>
  }
}
%alloc_0 = memref.alloc() {alignment = 64 : i64} : memref<32x32xf32>
scf.for %arg1 = %c0 to %c32 step %c1 {
  scf.for %arg2 = %c0 to %c32 step %c1 {
    %2 = memref.load %alloc[%arg1, %arg2] : memref<32x32xf32>
    memref.store %2, %alloc_0[%arg1, %arg2] : memref<32x32xf32>
  }
}
memref.dealloc %alloc : memref<32x32xf32>
scf.for %arg1 = %c0 to %c32 step %c1 {
  scf.for %arg2 = %c0 to %c32 step %c1 {
    scf.for %arg3 = %c0 to %c32 step %c1 {
      %2 = memref.load %cast[%arg1, %arg3] : memref<32x32xf32>
      %3 = memref.load %0[%arg3, %arg2] : memref<32x32xf32>
      %4 = memref.load %alloc_0[%arg1, %arg2] : memref<32x32xf32>
      %5 = arith.mulf %2, %3 : f32
      %6 = arith.addf %4, %5 : f32
      memref.store %6, %alloc_0[%arg1, %arg2] : memref<32x32xf32>
    }
  }
}
%alloc_1 = memref.alloc() {alignment = 64 : i64} : memref<32x32xf32>
scf.for %arg1 = %c0 to %c32 step %c1 {
  scf.for %arg2 = %c0 to %c32 step %c1 {
    %2 = memref.load %alloc_0[%arg1, %arg2] : memref<32x32xf32>
    %3 = memref.load %1[%arg2] : memref<32xf32>
    %4 = arith.addf %2, %3 : f32
    memref.store %4, %alloc_1[%arg1, %arg2] : memref<32x32xf32>
  }
}
\end{minted}
\end{minipage}\caption{Loop-level representation for \protect\mintinline{python}{torch.nn.Linear(32, 32)}
through \protect\mintinline{mlir}{torch-mlir}, \protect\mintinline{mlir}{linalg}, and \protect\mintinline{mlir}{scf}. The blue shading highlights the matrix-multiplication loop nest (lines 17-31)
amidst artifacts of the lowering process.}
\label{listing1}
\end{figure}
 The least abstract dialects at this level of abstraction are the
\mintinline{mlir}{arith} dialect, which models basic integer and
floating point mathematical operations, and the \mintinline{mlir}{vector}
dialect, a generic, retargetable, higher-order (i.e., multi-dimensional)
vector that carries semantically useful information for transformations
that enable targeting vector ISAs on concrete targets (e.g., AVX-512,
ARM SVE, etc.).

The dialects at this level of abstraction, especially \mintinline{mlir}{scf}
and \mintinline{mlir}{vector}, are where the real optimization work
occurs; transformations such as loop-unrolling, loop-fusion, loop-tiling
can have enormous impact on the runtime performance of any code \citep{10.1145/3178372.3179509},
but are especially important for numerics intensive code, such as
can be found to constitute the majority of kernels in a DNN. Furthermore,
explicit vectorization (rather than auto-vectorization) is critical
to achieving good performance of various compute-intensive kernels,
deployed to both CPUs and GPUs \citep{dickson2011importance}. Hence,
it's important to be able to efficiently and effectively manipulate
representations of DNNs at this level of abstraction, even more-so
than what MLIR currently enables.

\subsubsection{Low-level dialects (target-specific code generation)}

At the lowest level of abstraction, MLIR contains implementations
of dialects that can interface with hardware specific runtimes and
ISAs, such as \mintinline{mlir}{nvvm}, which models NVPTX instructions,
\mintinline{mlir}{spirv}, and \mintinline{mlir}{llvm}, which faithfully
models LLVM IR and therefore enables targeting all backends supported
by LLVM (including \texttt{x86\_64} and \texttt{aarch64}/\texttt{arm64}
CPU ISAs). The latter dialect includes support for managed runtimes
on top of ISAs (such as OpenMP, in combination with the \mintinline{mlir}{omp}
dialect) and coroutines (in combination with the \mintinline{mlir}{async}
dialect). These target-specific, code-generation focused, dialects
enable end-to-end compilation of MLIR programs (e.g., DNNs) to a variety
of execution environments, including single-core CPU, multi-core (threaded)
CPU, and GPU, including SoTA NVIDIA platforms but also lesser known
vendors that implement the SPIR-V standard (see Section \ref{sec:Demonstration-and-Evaluation}
for demonstrations of \texttt{nelli}'s end-to-end compilation features).

\subsection{eDSL construction in Python}

Given a host language, there are (invariably) several ways to implement
an embedded domain-specific language; the set of avenues available
is only circumscribed by the facilities of the host language and the
goals of the DSL designer. \texttt{nelli} is embedded in Python and
so we discuss two eDSL implementation approaches (including merits
and deficiencies) with Python as the host language. Indeed, each of
these two approaches was validated (i.e., implemented) over the course
of developing \texttt{nelli} and discarded in favor of the chosen
approach (see Section \ref{sec:Implementation}).

\subsubsection{Compiling}

The most straightforward approach to implementing an eDSL in any host
language (conceptually) is to build a compiler using that language
for (a subset of) that language. This involves static (ahead-of-time)
source analysis, including lexing, abstract syntax tree (AST) construction,
control-flow analysis, type inference, and code generation (for the
target language, MLIR or otherwise). Suffice it to say, this is a
monumental undertaking. Nonetheless, the undertaking has been undertaken,
in the context of Python and, specifically, numerics intensive programs,
many times to varying degrees of success \citep{behnel2010cython,nuitka,10.1145/3578360.3580275}.

The scope of such an undertaking is slightly improved by
the fact that Python provides, in its standard library, source lexing
(for Python source code), AST construction, and AST traversal utilities
(in the \mintinline{python}{ast} package). But, comparatively speaking,
these aspects of the undertaking are the least challenging\footnote{Indeed, there exist many lexers and parsers for Python \citep{zimmerman2022langcc,parr1995antlr}
implemented in other, more performant, languages, i.e., preferable
alternatives to the \mintinline{python}{ast} package, if one's goal
were to build a Python compiler.}; the principal challenges are control-flow analysis and type inference.
With respect to the latter, Python's highly permissive runtime and
``duck typing''\footnote{Python is believed to be ``dynamically typed'': this is a widely
held misconception. In fact, every value manipulated by the Python
runtime is a subclass of \mintinline{python}{<class object>}: \mintinline{python}{(1).|\_\_|class|\_\_|.|\_\_|bases|\_\_| == (<class object>)}.
Thus, method resolution (\textbf{which can be patched at runtime})
determines the effective type of a value: \mintinline{python}{(1).|\_\_|class|\_\_|.|\_\_|mro|\_\_| == (<class int>, <class object>)}. } paradigm requires a compiler to reckon with all mutations of an instantiated
object; any object instance can be made to quack like a duck at any
point in the execution of a Python program. More seriously (supposing
property mutations were prevented), Python does not have nested lexical
scopes below the level of a function body: for example, in the following
\begin{figure}[H]
\begin{minipage}[t]{0.37\textwidth}%
\begin{minted}[tabsize=4]{python}
def leaky(a):
	if a % 2 == 0:
		b = 5
		c = 3 * b
	elif a == 5:
		b = "5"
		c = "3" + b
	else:
		pass
	return c
\end{minted}
\end{minipage}\label{python-escape-analysis-1}
\end{figure}

\noindent the conditional actually ``yields'' two values (\mintinline{python}{b}
in addition to \mintinline{python}{c}) and the same is true for all
such regions (i.e., \mintinline{python}{for}s and \mintinline{python}{with}s),
i.e., they ``leak'' definitions and (possibly) grow the use-def
chains of identifiers in subsequent regions. In addition, irrelevant
of lexical scoping, the conditional actually yields union types (\mintinline{python}{b, c: int |$\mid$ |str |$\mid$ |None})
and hence the target language needs to support such union types. Currently,
MLIR does not support such union types\footnote{Certainly MLIR supports modeling union types but recall that the broader
goal is to translate Python to existing MLIR dialects, rather than
mapping Python to a novel dialect.}. 

\subsubsection{Tracing\label{subsec:Tracing}}

An alternative to ahead-of-time (AOT) compilation of a program is
just-in-time (JIT) compilation, and in particular, compilation of
only the ordered sequence of operations executed during some execution
the program; such a compiler is called a \emph{tracing} JIT, alluding
to the ``tracing'' of the execution path of the program. Several
such tracing JITs have been built for general purpose Python \citep{10.1145/1565824.1565827,lam2015numba,pyston,pyjion}.
A tracing JIT approach obviates the need to perform control-flow analysis
and type-inference, because both are fully reified at runtime. However,
a tracing JIT does not eliminate the need to parse a source representation
of the program, e.g., as in the case of Python, the bytecode representation.
Indeed, Numba \citep{lam2015numba}, Pyston \citep{pyston}, and Pyjion
\citep{pyjion} compile CPython virtual machine bytecode instructions
(as opposed to textual source) directly to (target) assembly language\footnote{All three projects employ a more generic JIT (LLVM for the former
two and the CoreCLR \citep{troelsen2017philosophy} for the latter)
for the ``last mile'' of code generation.}. It's important to emphasize that while, in principle, each of Numba,
Pyston, and Pyjion can be used to compile entire Python programs,
they are frequently used as eDSLs for accelerated implementations
of the numerics intensive portions of Python code, through their partial-compilation
APIs (\mintinline{python}{@njit} and \mintinline{python}{pyjion.enable()},
for Numba and Pyjion respectively). 

An alternative to JIT compiling Python at the bytecode level (i.e.,
handling all opcodes), especially relevant for eDSL construction,
is instrumenting (``hooking'') only a subset of operations in the
host language. For example, function calls and arithmetic operations.
The various Python DNN frameworks (PyTorch \citep{paszke2017automatic},
TensorFlow \citep{https://doi.org/10.48550/arxiv.1603.04467}, JAX
\citep{frostig2018compiling}) take this approach; by restricting
user programs to make calls to functions in their own namespaces and
by overloading various operators on proxy objects (see Listing \ref{operator-overloading}),
the eDSL can wholly own the means of production\footnote{Recall, a production is a rewrite rule specifying a symbol substitution
that can be recursively performed to generate new symbol sequences.
A finite set of productions $P$ is the main component in the specification
of a formal grammar (such as that of a programming language).}, and thereby perform source-to-source translation.
\begin{figure}
\centering{}\renewcommand\figurename{Listing.}%
\begin{minipage}[t]{0.7\textwidth}%
\begin{minted}[tabsize=4]{python}
class Tensor:
	def __add__(self, other: Tensor):
		emit(f"tensor.add %{self}, %{other}")
		...
	def __mul__(self, other: Tensor):
		emit(f"tensor.mult %{self} %{other}")
		...
	def __getitem__(self, item: tuple[int]):
		# indexed load
		emit(f"tensor.extract ${self}[{*item}]")
		...
	def __setitem__(self, key: tuple[int], value):
		# indexed store
		emit(f"tensor.insert ${value} into ${self}[{*item}]")
		...
\end{minted}
\end{minipage}\caption{Sketch of operator overloading on a proxy \protect\mintinline{python}{Tensor}
object for purposes of performing translation to the MLIR \protect\mintinline{mlir}{tensor}
dialect.}
\label{operator-overloading}
\end{figure}
While simple and effective, hooking function calls and operator overloading
suffers from an aesthetically displeasing deficiency: in a host language
(such as Python) where control-flow primitives such as \mintinline{python}{if}s
and \mintinline{python}{for}s cannot be instrumented, they must be
replaced (within the context of the eDSL) with explicit proxies (e.g.,
\mintinline{python}{tf.while|\_|loop} and \mintinline{python}{jax.lax.cond}).
More critically, existing such eDSLs suffer from a fundamental limitation
of the tracing approach: if host-language conditionals are allowed
in any capacity, then the path less traveled by the program will be
not captured by the eDSL. For example, in the following 
\begin{figure}[H]
\begin{minipage}[t]{0.4\textwidth}%
\begin{minted}[numbers=left,tabsize=4,highlightlines={10}]{python}
def single_path(x: Tensor, a: int):
	if a % 2 == 0:
		y = 2 * x
	else:
		y = 3 * x
	return y
\end{minted}
\end{minipage}
\end{figure}

\noindent Despite being able to effectively capture all arithmetic
operations on a \mintinline{python}{Tensor}, no tracing eDSL can
capture both arms of the conditional. \texttt{nelli} addresses this
limitation.

\section{Design and implementation of \texttt{nelli}\label{sec:Implementation}}

The primary design goal of \texttt{nelli} is to be \emph{easy to use}
and \emph{simple to understand}, while remaining faithful to the semantics
of MLIR. By semantics of MLIR, we mean that dialects as rendered in
\texttt{nelli} (i.e., names and uses of operations) should reflect
as closely as possible their rendering in MLIR IR. Note, we draw a
subtle distinction between easy and simple: easy to use implies that
it should work (generate MLIR IR) with very little fanfare while simple
to understand means studying the implementation should reward a modicum
of effort (without requiring an inordinate investment). Addressing
the former, much effort on our part has been invested in packaging
\texttt{nelli} for distribution (it can be directly \texttt{pip install}ed
without compiling LLVM/MLIR). Further, in order to reduce the barrier to reuse of existing code, \texttt{nelli} is also extensible (in and of itself) and exposes MLIR in an extensible way.

Addressing the latter precludes various
metaprogramming techniques, such as wholesale source rewriting and
Python \mintinline{python}{metaclass} programming. Additionally,
it precludes the use of dynamic scoping (using \mintinline{python}{contextvars})
to implement patterns such as stacks of monadic interpreters \citep{kiselyov2012typed,10.1145/3158140}.
\texttt{nelli} uses three techniques to accomplish the stated design
goals: operator overloading, trivial source rewriting, and bytecode
rewriting. We discuss each in turn (effectively, in order of increasing
complexity). We also discuss how \texttt{nelli} addresses extensibility.

\subsection{Upstream manicuring and operator overloading}

\begin{figure}
\centering{}\renewcommand\figurename{Listing.}%
\begin{minipage}[t]{0.6\textwidth}%
\begin{minted}[tabsize=4]{python}
with Context() as ctx:
	with Location.unknown(context=ctx) as loc:
		index_type = IndexType.get()
		f = func.FuncOp("simple_for", ([], []))
		with InsertionPoint(f.add_entry_block()):
		    lb = arith.ConstantOp.create_index(0)
		    ub = arith.ConstantOp.create_index(42)
		    step = arith.ConstantOp.create_index(2)
			three = arith.ConstantOp.create_index(3)
		    loop = scf.ForOp(lb, ub, step, iter_args)
		    with InsertionPoint(loop.body):
				three_i = arith.MulIOp(
					three, 
					loop.induction_variable
				)
		        scf.YieldOp([])
		    func.ReturnOp([])
\end{minted}
\end{minipage}\enskip{}%
\begin{minipage}[t]{0.35\textwidth}%
\begin{minted}[tabsize=4]{python}
@mlir_func
def simple_for():
    for i in range(0, 42, 2):
		two_i = 3 * i
\end{minted}
\end{minipage}\caption{Instantiating \protect\mintinline{mlir}{func.func} with a \protect\mintinline{mlir}{scf.for}
using the upstream MLIR Python bindings compared with specifying the same program using \texttt{nelli}.}
\label{mlirfuncop}
\end{figure}

MLIR, irrelevant of \texttt{nelli}, procedurally generates Python
bindings for functionality related to emitting MLIR IR. This procedural
generation is made possible by virtue of the fact that almost all
operations, in all MLIR dialects, are defined using ODS (see Section
\ref{subsec:MLIR}). Nonetheless, convenient (and robust) as these
existing bindings might be, they are quite verbose, requiring specifying
most attributes of operations explicitly; see Listing \ref{mlirfuncop}
for an example. Thus, some of the work of \texttt{nelli} involves
normalizing the upstream APIs; in particular we implement operator
overloading for various arithmetic operations on values that are results
of \mintinline{mlir}{arith} operations (see Listing \ref{mlirfuncop-1}),
as well indexing and slicing on results of \mintinline{mlir}{memref}
and \mintinline{mlir}{tensor} operations (see Listing \ref{nelli_for_to_affine_for}).
Additionally we overload Python parameter annotations to implement
a minimal form of Hindley-Milner\footnote{In reality, MLIR performs the type inference, \texttt{nelli} simply
requires fully type-annotated function parameters.}, as well as instantiating \mintinline{mlir}{func}s with typed parameters.
Finally, we use Python \mintinline{python}{class} namespaces as models
for \mintinline{mlir}{module}s, including nested \mintinline{mlir}{gpu.module}s
(see Listing \ref{gpumodules}).

\begin{figure}
\centering{}\renewcommand\figurename{Listing.}%
\begin{minipage}[t]{0.45\textwidth}%
\begin{minted}[tabsize=4]{python}
class MyClass1(GPUModule):
    def kernel(
        self,
        A: MemRef[(M, N), F32],
        B: MemRef[(N, K), F32],
        C: MemRef[(M, K), F32],
    ):
        x = block_id_x()
        y = block_id_y()
        a = A[x, y]
        b = B[x, y]
        C[x, y] = a * b
        return

m = MyClass1(
    func_attributes={
        "spirv.entry_point_abi": 
			spirv.entry_point_abi(
            	workgroup_size=[1, 1, 1]
			),
    }
)

@mlir_func
def main(
    A: MemRef[(M, N), F32],
    B: MemRef[(N, K), F32],
    C: MemRef[(M, K), F32],
):
    m.kernel(A, B, C, 
		grid_size=[4, 4, 1], 
		block_size=[1, 1, 1]
	)
\end{minted}
\end{minipage}\enskip{}%
\begin{minipage}[t]{0.52\textwidth}%
\begin{minted}[tabsize=4]{mlir}
module attributes {gpu.container_module} {
  gpu.module @MyClass1 {
    gpu.func @kernel(
		%A: memref<4x16xf32>, 
		%B: memref<16x8xf32>, 
		%C: memref<4x8xf32>) 
	kernel attributes {
		spirv.entry_point_abi = 
			#spirv.entry_point_abi<
				workgroup_size = [1, 1, 1]
			>
	} {
      %x = gpu.block_id  x
      %y = gpu.block_id  y
      %a = memref.load %A[%x, %y] : ...
      %b = memref.load %B[%X, %Y] : ...
      %c = arith.mulf %a, %b : f32
      memref.store %C, %C[%0, %1] : ...
      gpu.return
    }
  }
  func.func @main(
	%A: memref<4x16xf32>, 
	%B: memref<16x8xf32>, 
	%C: memref<4x8xf32>) {
    %c4 = arith.constant 4 : index
    %c1 = arith.constant 1 : index
    gpu.launch_func async 
	  @MyClass1::@kernel 
	  blocks in (%c4, %c4, %c1) 
	  threads in (%c1, %c1, %c1) 
	  args(
		%A : memref<4x16xf32>, 
		%B : memref<16x8xf32>, 
		%C : memref<4x8xf32>
	  )
    return
  }
}
\end{minted}
\end{minipage}\caption{Overloading \protect\mintinline{python}{class}es to support nested
\protect\mintinline{mlir}{gpu.module}s.}
\label{gpumodules}
\end{figure}

\begin{figure}
\centering{}\renewcommand\figurename{Listing.}%
\begin{minipage}[t]{0.4\textwidth}%
\begin{minted}[tabsize=4]{python}
one = arith.constant(1.0)
two = arith.constant(2.0)
three = one + two
\end{minted}
\end{minipage}\enskip{}%
\begin{minipage}[t]{0.45\textwidth}%
\begin{minted}[tabsize=4]{mlir}
%one = arith.constant 1.0e+00 : f32 
%two = arith.constant 2.0e+00 : f32 
%three = arith.addf %one, %two : f32
\end{minted}
\end{minipage}\caption{Operator overloading of results of \protect\mintinline{mlir}{arith}
operations.}
\label{mlirfuncop-1}
\end{figure}

\subsection{Trivially rewriting the AST}

It's important to understand how the upstream MLIR Python bindings
function (as a reflection of how MLIR functions). Consider the instantiation
of \mintinline{mlir}{scf.for} in Listing \ref{mlirfuncop}; operations
to be inserted into the body of the \mintinline{mlir}{scf.for} must
have their \mintinline{python}{InsertionPoint}s set to (somewhere
in) the body of the \mintinline{mlir}{scf.for}. Thus, the Python
bindings corresponding to those operations (i.e., \mintinline{python}{arith.MulIOp})
must be executed within the context of \mintinline{python}{with InsertionPoint(loop.body)}.
Eliminating the indentation due to the \mintinline{python}{with}
(which indicates a nested scope where none exists) is worthwhile.
One trivial way to accomplish this is to explicitly \mintinline{python}{|\_\_|enter|\_\_|}
and \mintinline{python}{|\_\_|exit|\_\_|} the \mintinline{python}{InsertionPoint(loop.body)}
context manager; see Listing \ref{mlirfuncop-2}.
\begin{figure}
\centering{}\renewcommand\figurename{Listing.}%
\begin{minipage}[t]{0.35\textwidth}%
\begin{minted}[tabsize=4]{python}
@mlir_func(rewrite_ast_=True)
def simple_for():
    for i in range(0, 42, 2):
		two_i = 3 * i
\end{minted}
\end{minipage}\quad{}\raisebox{-3em}{$\Rightarrow$}\quad{}%
\begin{minipage}[t]{0.4\textwidth}%
\begin{minted}[tabsize=4]{python}
@mlir_func
def simple_for():
    for i in scf_range(0, 42, 2):
		two_i = 3 * i
        scf_endfor()
\end{minted}
\end{minipage}\caption{Trivially rewriting user functions in order to explicitly manage context
managers for MLIR operations with regions; the \protect\mintinline{python}{scf|\_|range}
(in addition to instantiating the \protect\mintinline{mlir}{scf.for})
triggers \protect\mintinline{python}{|\_\_|enter|\_\_|} (on a thread-local
handle to a context manager) and the \protect\mintinline{python}{scf|\_|range}
triggers \protect\mintinline{python}{|\_\_|exit|\_\_|}.}
\label{mlirfuncop-2}
\end{figure}
But requiring the user to explicitly indicate the end of the \mintinline{python}{for}
loop transforms Pythonic \mintinline{python}{for} loops to Pascal-style
\mintinline{pascal}{for} loops. Thus, \texttt{nelli} rewrites user
functions (at the AST level) and automatically inserts such opening
and closing calls for all \mintinline{python}{for}s and \mintinline{python}{if}s
(see Listing \ref{mlirfuncop-2-1}) 
\begin{figure}
\centering{}\renewcommand\figurename{Listing.}%
\begin{minipage}[t]{0.35\textwidth}%
\begin{minted}[tabsize=4]{python}
@mlir_func(rewrite_ast_=False)
def ifs(M: F64, N: F64):
    one = constant(1.0)
    if scf_if(M < N):
        one = constant(1.0)
        scf_endif_branch()
    else:
        scf_else()
        two = constant(2.0)
        scf_endif_branch()
        scf_endif()
\end{minted}
\end{minipage}\quad{}\raisebox{-8em}{$\Rightarrow$}\quad{}%
\begin{minipage}[t]{0.4\textwidth}%
\begin{minted}[tabsize=4]{python}
@mlir_func(rewrite_ast_=True)
def ifs(M: F64, N: F64):
    one = constant(1.0)
    if M < N:
        one = constant(1.0)
    else:
        two = constant(2.0)
\end{minted}
\end{minipage}\caption{AST rewriting of conditionals for manual (but implicit) management
of context managers for lowering to \protect\mintinline{mlir}{scf.if}. }
\label{mlirfuncop-2-1}
\end{figure}
. Note, since we rewrite the AST (not the source itself), we are able
to patch line numbers for all nodes to reflect original source locations
and thus all Python IDE, error-reporting, and debugging infrastructure
is undeterred i.e., users are able to set breakpoints in functions
and inspect objects just the same as for any Python code\footnote{This is emphatically not the case for eDSLs like Numba and Pyjion
which compile and execute Python using, effectively, their own bytecode
interpreters.}. 

\subsection{Trivially rewriting bytecode}

Rewriting the source AST enables mapping Python control-flow primitives
to various MLIR control-flow operations except for one caveat: as
mentioned in section \ref{subsec:Tracing}, relying on runtime execution
of Python code (and hooks) to capture programs precludes faithful
capture of conditionals. For example, irrespective of arbitrary AST
transformations, only one arm of the conditional in Listing \ref{mlirfuncop-2-1}
can be traced. Although it's debatable whether multi-arm conditionals
are crucial (\mintinline{mlir}{scf.if} does support an \mintinline{mlir}{scf.else}
branch), it would be a strange language that supported only single-arm
conditionals. 

The resolution to the conundrum of the multi-arm conditional lies in rewriting the program
on a deeper level than the AST; recall Python programs are compiled
(by the CPython implementation) to bytecode instructions. 
\begin{figure}
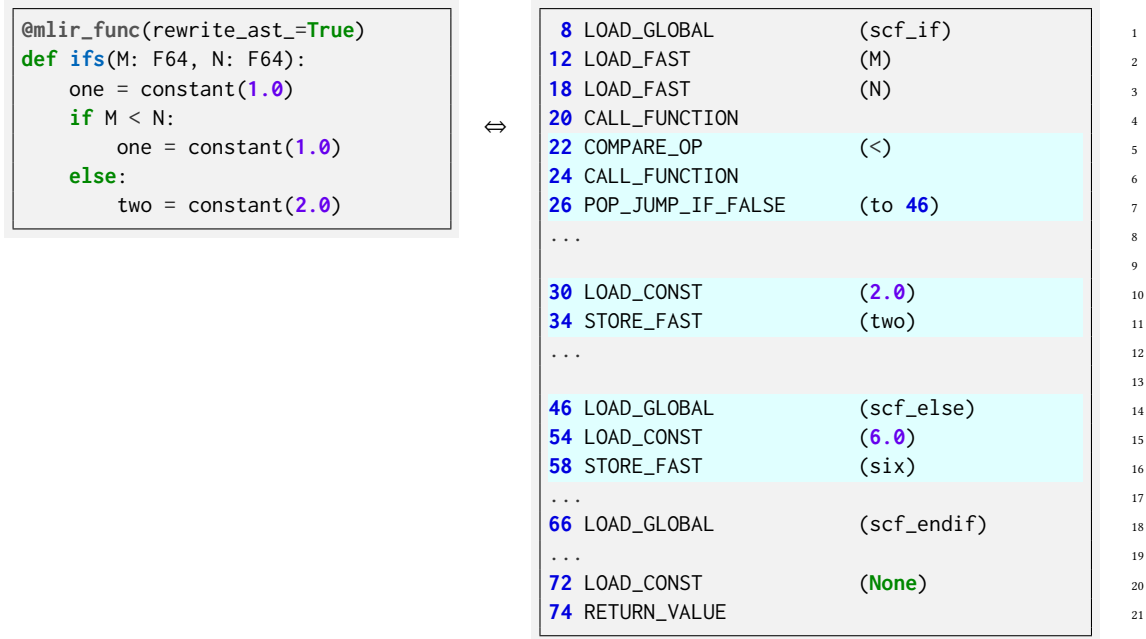

\centering{}\renewcommand\figurename{Listing.}%
\begin{minipage}[t]{0.4\textwidth}%
\begin{minted}[tabsize=4]{python}
@mlir_func(rewrite_ast_=True)
def ifs(M: F64, N: F64):
    one = constant(1.0)
    if M < N:
        one = constant(1.0)
    else:
        two = constant(2.0)
\end{minted}
\end{minipage}\quad{}\raisebox{-6em}{$\Leftrightarrow$}\quad{}%
\begin{minipage}[t]{0.5\textwidth}%
\begin{minted}[numbers=right,tabsize=4,highlightlines={5-7,10-11,14-16}]{python}
 8 LOAD_GLOBAL            (scf_if)
12 LOAD_FAST              (M)
18 LOAD_FAST              (N)
20 CALL_FUNCTION          
22 COMPARE_OP             (<)
24 CALL_FUNCTION          
26 POP_JUMP_IF_FALSE      (to 46)
...
|\tikzmark{sumtop}|
30 LOAD_CONST             (2.0) 
34 STORE_FAST             (two)|\tikzmark{sumbottom}| 		
...			
|\tikzmark{maintop}|
46 LOAD_GLOBAL            (scf_else)
54 LOAD_CONST             (6.0) 
58 STORE_FAST             (six)|\tikzmark{mainbottom}| 		
...
66 LOAD_GLOBAL            (scf_endif)
...
72 LOAD_CONST             (None)
74 RETURN_VALUE
\end{minted}
\end{minipage}\caption{CPython bytecode instructions corresponding to multi-arm \protect\mintinline{python}{if};
note the \protect\mintinline{python}{POP|\_|JUMP|\_|IF|\_|FALSE}
that executes a jump to the else branch (lines 14-16) if the condition
(\protect\mintinline{python}{COMPARE|\_|OP}) evaluates to \protect\mintinline{python}{False}
(whereas, otherwise the true branch, lines 10-11, is executed).}
\label{mlirfuncop-2-1-1}
\end{figure}
 The CPython implementation of Python is a stack-based virtual machine
\citep{ike2015inside} that implements conditionals like many other
virtual machines: using jump instructions (see Listing \ref{mlirfuncop-2-1-1}).
Thus, the solution is to simply rewrite the bytecode of the user's
function and remove those jumps\footnote{In fact, the jump instruction is replaced by a \mintinline{python}{NOP}
(no-op) instruction in order to prevent invalidating stack size calculations.}, thereby forcing the CPython interpreter to execute all instructions
in all arms of the conditional. It's important to emphasize that this
transformation is reasonable given the stated goals of \texttt{nelli}:
the eDSL program isn't computing on data and has no intended side-effects
other than to emit MLIR IR. Thus program capture, rather than evaluation,
permits (and encourages) us to fundamentally alter the semantics of
conditionals in this way; certainly, under different circumstances,
such a transformation would be wholly nonsensical.

\subsection{Extensibility}

Currently MLIR is extensible to a limited extent\footnote{The state of affairs is steadily improving; between starting and finishing this manuscript, the authors contributed three substantive improvements.}
\texttt{nelli} addresses extensibility in four ways: the first and second being exercises of existing (but infrequently employed) MLIR APIs, and with the remaining two being novel (relative to MLIR):
\begin{enumerate}
    \item \texttt{nelli} uses the \mintinline{python}{_site_initialize} to register (in-tree) dialects at load-time;
    \item \texttt{nelli} subclasses existing \mintinline{python}{ir.OpView} classes thereby extending their functionality despite being out-of-tree;
    \item \texttt{nelli} is built with exported symbols, thus enabling users to extend the various C++ utility classes that comprise the upstream bindings (without recompiling MLIR);
    \item \texttt{nelli} implements AST walking functionality (akin to Python's \mintinline{python}{ast.NodeTransformer}) which, along with upstream improvements contributed by the authors, enables writing simple IR rewrites wholly in Python.
\end{enumerate}
Extension points (1) and (2) exercise existing MLIR Python bindings APIs in unconventional ways to demonstrate (1) extending the set of registered (available at runtime) dialects and (2) wrapping/polishing existing \mintinline{python}{ir.OpView} APIs (constructors, getters, setters) without repackaging the bindings. Extension points (3) and (4) are more nuanced.

By default, MLIR Python bindings are built with "hidden" symbols (i.e., \mintinline{cpp}{-fvisibility=hidden})\footnote{This is actually a design choice made by \mintinline{cpp}{pybind11} \citep{pybind11} rather than MLIR.}, making those symbols unavailable for symbol resolution of subsequently loaded libraries. 
This has the effect that none of the bindings' utility classes can be extended without recompiling.
\texttt{nelli} exports these symbols (i.e., annotates them with \mintinline{cpp}{__attribute__((visibility("default")))}), thus downstream projects can \mintinline{bash}{pip install nelli} immediately extend bindings for existing MLIR dialects, operations, types, and attributes.
With respect to the final extension point (writing simple IR rewrites wholly in Python), recall that the conventional method for transforming/rewriting IR is by building a \mintinline{cpp}{RewritePattern} using the C++ API.
While the MLIR C++ API is robust and well-designed, like all such APIs, it is rigid and demanding; that is to say, it is not suited for experimentation and quick iteration.
By contrast the Python AST utilities (under the standard library package \mintinline{python}{ast}) present a lightweight API that enables building small (but effective) rewrites of the Python AST.
Taking this user experience as our inspiration, we implement similar functionality by generating AST visitors procedurally from the upstream bindings; one \mintinline{python}{DialectVisitor} for each registered MLIR dialect, including out-of-tree dialects.
Using these \mintinline{python}{DialectVisitor}s, along with the recently contributed \mintinline{python}{replace_all_uses_with} upstream API, we are able to build small but non-trivial IR transformations on MLIR \mintinline{python}{ir.Module}s just as one does using \mintinline{python}{ast.NodeTransformer}.

\section{Demonstration and evaluation}\label{sec:Demonstration-and-Evaluation}

We demonstrate the capabilities of \texttt{nelli} with three small exercises:

\begin{enumerate}
    \item an end-to-end (including execution) GPU example of a batched, multi-channel, 2D convolution that lowers to both NVIDIA devices and Vulkan supporting devices, such as the VideoCore VI 3D found in the Raspberry Pi 4 Model B and the Apple Neural Engine (found in the Apple M1 series of laptops);
    \item an end-to-end that lowers the same kernel to the managed OpenMP runtime;
    \item the previous two examples integrated with a black-box, gradient-free, optimizer \citep{nevergrad} for searching the space of possible optimizing transformations.
\end{enumerate}

\subsection{End-to-end GPU}

\begin{figure}
\centering{}\renewcommand\figurename{Listing.}%
\begin{minipage}[t]{0.50\textwidth}%
\begin{minted}[tabsize=4]{python}
@mlir_func(range_ctor=scf_for)
def conv2d_nchw_fchw(
  input: MemRef[(N, CI, HI, WI), F32],
  kernel: MemRef[(CO, CI, K, K), F32],
  output: MemRef[(N, CO, HO, WO), F32],
):
  for n, co, ho, wo in parallel(
    (0, 0, 0, 0), (N, CO, HO, WO)
  ):
    for ci in range(0, CI):
      for ki in range(0, K):
        for kj in range(0, K):
          ii = ho + ki
          jj = wo + kj
          inp = input[n, ci, ii, jj]
          ker = kernel[co, ci, ki, kj]
          output[n, co, ho, wo] += inp * ker
\end{minted}
\end{minipage}
\hfill{}%
\begin{minipage}[t]{0.47\textwidth}%
\begin{minted}[tabsize=4,numbers=right,highlightlines={13,14},numbersep=5pt]{python}
pipeline=Pipeline()
    .FUNC()
      .gpu_map_parallel_loops()
    .CNUF()
    .convert_parallel_loops_to_gpu()
    .FUNC()
      .lower_affine()
      .convert_scf_to_cf()
      .gpu_kernel_outlining()
    .CNUF()
    .GPU()
      .strip_debuginfo()
      .convert_gpu_to_nvvm()
      .gpu_to_cubin(chip="sm_75")
    .UPG()
    .gpu_to_llvm()
\end{minted}
\end{minipage}
\caption{Standard representation of a 2D-NCHW convolution parallelized across output elements and corresponding pass pipeline. Note, we freely interleave \mintinline{mlir}{scf.parallel} and \mintinline{mlir}{scf.for} (i.e., \mintinline{python}{range}). Note also that only lines 13, 14 of the pass pipeline are specific to NVIDIA devices.}
\label{2dnchwconv}
\end{figure}

We implement a standard 2D-NCHW convolution and parallelize across output elements; see Listing \ref{2dnchwconv}. 
Note, we set the default range to be mapped to \mintinline{mlir}{scf.for} (\mintinline{python}{range_ctor=scf_for}) but explicitly map the outermost four loops to a \mintinline{mlir}{scf.parallel} thereby specifying that each output element (\mintinline{python}{n, co, ho, wo}) should be computed in parallel.
The pass pipeline (cf. Listing \ref{2dnchwconv}) that accompanies the kernel is specialized for NVIDIA devices but only in the final (hardware-specific) passes. 
The higher-level passes effectively perform two functions: \mintinline{python}{gpu_map_parallel_loops} assigns nested \mintinline{mlir}{scf.parallel} loops to the corresponding level of the GPU workgroup hierarchy (grid, block, and thread) and \mintinline{python}{gpu_kernel_outlining} outlines GPU kernel code so that it can be separately compiled and serialized (\mintinline{python}{gpu_to_cubin(chip="sm_75")}).
Note that while this basic example only has one \mintinline{mlir}{scf.parallel} and is thus mapped only to blocks, further transformations (such as tiling) can introduce nested \mintinline{mlir}{scf.parallel}s, thereby inducing a distribution across both blocks and threads.

The presented code lowers fully to both NVIDIA and Vulkan targets almost unaltered\footnote{Vulkan does not support 4D buffers so in practice inputs have to be packed along the batch and channel dimensions.} and therefore this kernel successfully executes on all three of the aforementioned hardware platforms (NVIDIA 3080Ti, Apple Neural Engine, and VideoCore VI). In fact, since MLIR provides utilities for mapping NumPy arrays to the various GPU runtime native buffers (using a buffer descriptor called \mintinline{cpp}{StridedMemRef}), and utilities for interfacing with the hardware runtimes (CUDA runtime and Vulkan runtime), all end-to-end experiments can be executed without ever leaving the comfort of \texttt{nelli}.

\subsection{End-to-end OpenMP}

Since the aforementioned 2D-NCHW convolution kernel is implemented in terms of the \mintinline{mlir}{scf} dialect, which, as the name suggests, models programs in terms of abstract but structured control-flow operations, the same 2D-NCHW convolution kernel can be lowered to the OpenMP runtime with just the flip of a switch; using \texttt{nelli}'s \mintinline{python}{Pipeline}, it's just a matter of substituting \mintinline{python}{convert_scf_to_openmp} for \mintinline{python}{convert_parallel_loops_to_gpu}.
This wraps the loop nest in a \mintinline{mlir}{omp.parallel} context and implements the \mintinline{mlir}{scf.parallel} loop as a \mintinline{mlir}{omp.wsloop}, i.e., worksharing loop.
Just as with the end-to-end GPU implementation, thanks to LLVM's support for OpenMP, all end-to-end experiments can be executed without ever leaving the comfort of \texttt{nelli}.

\subsection{Derivative-free optimization}

\begin{figure}
  \begin{centering}
  \includegraphics[scale=0.5]{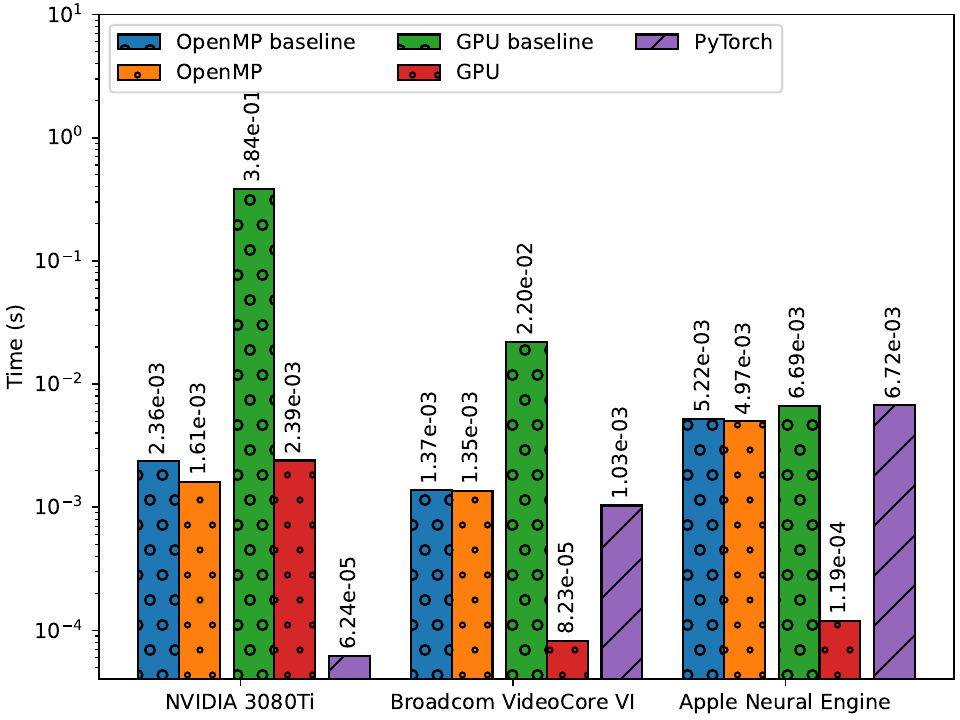}\caption{Nevergrad optimization for tiling and inner-loop unrolling of a 2D-NCHW convolution kernel.}\label{nevergrad}
  \par\end{centering}
\end{figure}

Algorithmic correctness is a necessary but not sufficient condition for achieving high performance implementations of numerically intensive kernels; due to the variety of hardware platforms, fine-tuning of the implementation for each platform is critical \citep{li2009note}. 
In many cases, where a formal cost model of the hardware platform isn't available, a directed search of the program transformation space cannot be realized.
In such instances, black-box (gradient-free) optimization techniques can be employed.
To demonstrate the value of having seamless interoperability with the Python ecosystem, we connect \texttt{nelli} to \texttt{Nevergrad} \citep{nevergrad}, a Python package for gradient-free optimization.
We experiment with applying two loop nest transformations: tiling and unrolling \citep{CARDOSO2017137}, using the MLIR pass \mintinline{python}{scf_parallel_loop_tiling} and a loop-unrolling operation (\mintinline{python}{loop_ext.LoopUnrollOp}).
Both transformations are parameterized (by tile sizes and unroll factor, respectively) and it is this parameter space that can be searched over to find the optimal kernel for each hardware platform.

Thus, we set \texttt{Nevergrad} loose on our kernel on three different platforms: a workstation with a NVIDIA 3080Ti GPU, an Apple M1 MacBook Pro with a Neural Engine GPU, and a Raspberry Pi 4 Model B with Broadcom VideoCore VI GPU.
On each platform we apply the 2D-NCHW convolution on an input with shape $\left(N, C, H, W\right) := \left(1, 1, 1280, 1280\right)$, using a $\left(C_i, C_o, K\right) := \left(1, 3, 3\right)$ filter (where $C_i, C_o$ are channels-in, channels-out, respectively), and search the space of possible tile sizes and loop-unroll factors.
In addition, just for fun, we apply loop-unrolling to the OpenMP implementation.
As well, we compare it to the PyTorch implementation of the same kernel.
Figure \ref{nevergrad} shows the results of the experiment.
A few noteworthy observations:
\begin{enumerate}
  \item for this small kernel, OpenMP is fairly performant but loop-unrolling has almost no effect on the implementation because the worksharing loop distribution already distributes work maximally across all available cores;
  \item tiling and loop-unrolling is hugely important for achieving good performance on GPUs;
  \item PyTorch has highly optimized implementations of kernels for common platforms (such as NVIDIA) but otherwise can be outperformed by fine-tuning for a user's specific hardware configuration.
\end{enumerate}
Indeed, each of these conclusions is well-known and understood and thus we observe that enabling users to easily, and transparently, perform this kind of fine-tuning, through a representation like MLIR and an interface like \texttt{nelli}, is a boon to mankind.

\section{Related Work\label{sec:Related-Work}}

There are several projects in the MLIR ecosystem that aim to support a Python frontend for some MLIR dialect (or collection of dialects) but as far as we're aware (i.e., at the time of writing) none that aim to expose \emph{all} builtin dialects through a Python frontend.
Most famous amongst these are JAX~\citep{jax2018github}, which provides a NumPy-conformant interface to the \mintinline{mlir}{hlo}~\citep{stablehlo} family of dialects, and TensorFlow.
Each of these effectively provides a very high-level, Python-embedded, DSL for machine learning (specifically neural network) operations implemented against their respective dialects.
Note, while superficially the Torch-MLIR project~\citep{torchmlir} resembles JAX and TensorFlow, in that it transforms Python to an MLIR dialect (\mintinline{mlir}{torch-mlir}), the Python frontend is in fact provided by PyTorch itself\footnote{Torch-MLIR operates an the intermediate representations exported by PyTorch rather than directly on Python source.}, we do not count it amongst this group of projects.
In this category there is also the interesting \texttt{numba-mlir} project, a MLIR-based Numba backend, where, Numba~\citep{10.1145/2833157.2833162} translates from NumPy-specific Python code to lower-level representations (Numba IR and then LLVM IR) by analyzing the CPython bytecode instructions\footnote{CPython is a stack-based virtual machine with its own assembly/bytecode instructions (see \url{https://docs.python.org/3/library/dis.html}).} the Python compiles to.
\texttt{numba-mlir} lowers to several high-level (higher than LLVM IR) domain-specific dialects (\mintinline{mlir}{ntensor}, \mintinline{mlir}{plier}) by recovering Regionalized Value State Dependence Graphs~\citep{10.1145/3391902} from the Numba IR.
Finally, there is the Pylir project, implemented on top of MLIR, which aims to be an optimizing, ahead-of-time, compiler for all of Python.
Thus, Pylir's goal is not to be a frontend for MLIR in and of itself but to compile Python code to native executables; it accomplishes this impressive feat by parsing Python to its own dialects.

Alternatively, there exist projects that approach the problem orthogonally - they aim to provide a Python interface to a MLIR-like framework but not MLIR itself. 
The projects in this category generally aim to be completely independent of upstream MLIR and thus parse MLIR-native IR into proprietary ASTs and manipulate those ASTs in various ways (transforming, serializing, etc.).
For example, pyMLIR~\citep{pymlir} implements a LALR(1) grammar (extracted from the upstream MLIR documentation) and parser. 
pyMLIR's parser generates an AST representation that further implements Python's \mintinline{python}{ast.NodeTransformer} interface, thus enabling various AST transformations.
The resulting ASTs can be once again serialized to conformant MLIR.
Note, \texttt{nelli}'s AST transformation functionality is inspired by pyMLIR's, but instead of operating on a proprietary AST representation, \texttt{nelli}'s operates on the canonical MLIR AST.
On the other hand, xDSL~\citep{brownxdsl} is a Python-native compiler framework influenced by MLIR but not coupled directly to MLIR; xDSL emitted IR is validated against MLIR but only as part of its continuous integration process.
These projects are all very interesting and impressive accomplishments but they are orthogonal to our goals, i.e., providing a Python frontend to MLIR itself, rather than an arbitrary compiler framework (irrespective of how featureful or MLIR-like that framework might be).

\section{Conclusion\label{sec:Conclusion}}

We described \texttt{nelli}, a lightweight, open source, Python frontend for the Multi-Level Intermediate Representation compiler infrastructure.
\texttt{nelli} aims to make MLIR more accessible by providing a Pythonic syntax for the various MLIR dialects while remaining faithful to MLIR's semantics.
\texttt{nelli} uses operator overloading, AST rewriting, and bytecode rewriting to map Python control flow primitives, like conditionals and loops, to MLIR control flow operations, amongst other design choices.
\texttt{nelli} is designed to be simple to use and understand. It performs minimal static analysis and thus incurs minimal complexity in perform the translation to MLIR, compared to existing frontends. 
As a Python eDSL, it interoperates with existing Python tooling is fully extensible, in terms of dialects, operations, types, and attributes supported.
Further, we demonstrated the utility of \texttt{nelli} by showing end-to-end compilation of an example kernel for different hardware platforms, including integration with a derivative-free optimization library to automatically optimize for those platforms.
In summary, \texttt{nelli} provides an easy way to interface with MLIR and manipulate intermediate representations directly, avoiding the complexities and artifacts of lowering from high-level frameworks. This enables more flexible program analysis and transformation compared to those existing MLIR frontends.

\bibliographystyle{plain}
\bibliography{nelli}

\begin{thebibliography}{10}

\bibitem{nvidiasass}
Ptx and sass assembly debugging.
\newblock
  \url{https://docs.nvidia.com/gameworks/content/developertools/desktop/ptx_sass_assembly_debugging.htm},
  2015.
\newblock [Online; accessed 26-March-2023].

\bibitem{pyston}
Pyston.
\newblock \url{https://github.com/pyston/pyston}, 2023.
\newblock [Online; accessed 26-March-2023].

\bibitem{stablehlo}
{StableHLO}: Backward compatible ml compute opset inspired by hlo/mhlo, 2023.

\bibitem{torchmlir}
{Torch-MLIR}: The torch-mlir project aims to provide first class support from
  the pytorch ecosystem to the mlir ecosystem, 2023.

\bibitem{https://doi.org/10.48550/arxiv.1603.04467}
Martín Abadi, Ashish Agarwal, Paul Barham, Eugene Brevdo, Zhifeng Chen, Craig
  Citro, Greg~S. Corrado, Andy Davis, Jeffrey Dean, Matthieu Devin, Sanjay
  Ghemawat, Ian Goodfellow, Andrew Harp, Geoffrey Irving, Michael Isard,
  Yangqing Jia, Rafal Jozefowicz, Lukasz Kaiser, Manjunath Kudlur, Josh
  Levenberg, Dan Mane, Rajat Monga, Sherry Moore, Derek Murray, Chris Olah,
  Mike Schuster, Jonathon Shlens, Benoit Steiner, Ilya Sutskever, Kunal Talwar,
  Paul Tucker, Vincent Vanhoucke, Vijay Vasudevan, Fernanda Viegas, Oriol
  Vinyals, Pete Warden, Martin Wattenberg, Martin Wicke, Yuan Yu, and Xiaoqiang
  Zheng.
\newblock Tensorflow: Large-scale machine learning on heterogeneous distributed
  systems, 2016.

\bibitem{10.1145/3158140}
Nada Amin and Tiark Rompf.
\newblock Collapsing towers of interpreters.
\newblock 2(POPL), dec 2017.

\bibitem{pyjion}
{Anthony Shaw}.
\newblock Nuitka the python compiler.
\newblock \url{https://pyjion.readthedocs.io/en/latest/}, 2023.
\newblock [Online; accessed 26-March-2023].

\bibitem{behnel2010cython}
Stefan Behnel, Robert Bradshaw, Craig Citro, Lisandro Dalcin, Dag~Sverre
  Seljebotn, and Kurt Smith.
\newblock Cython: The best of both worlds.
\newblock {\em Computing in Science \& Engineering}, 13(2):31--39, 2010.

\bibitem{blockhaus2022framework}
Paul Blockhaus and Ing~David Broneske.
\newblock {\em A Framework for Adaptive Reprogramming Using a JIT-Compiled
  Domain Specific Language for Query Execution}.
\newblock PhD thesis, Master's thesis. Ottovon-Guericke University Magdeburg,
  2022.

\bibitem{10.1145/1565824.1565827}
Carl~Friedrich Bolz, Antonio Cuni, Maciej Fijalkowski, and Armin Rigo.
\newblock Tracing the meta-level: Pypy's tracing jit compiler.
\newblock In {\em Proceedings of the 4th Workshop on the Implementation,
  Compilation, Optimization of Object-Oriented Languages and Programming
  Systems}, ICOOOLPS '09, pages 18--25, New York, NY, USA, 2009. Association
  for Computing Machinery.

\bibitem{jax2018github}
James Bradbury, Roy Frostig, Peter Hawkins, Matthew~James Johnson, Chris Leary,
  Dougal Maclaurin, George Necula, Adam Paszke, Jake Vander{P}las, Skye
  Wanderman-{M}ilne, and Qiao Zhang.
\newblock {JAX}: composable transformations of {P}ython+{N}um{P}y programs,
  2018.

\bibitem{brownxdsl}
Nick Brown, Tobias Grosser, Mathieu Fehr, Michel Steuwer, and Paul Kelly.
\newblock xdsl: A common compiler ecosystem for domain specific languages.

\bibitem{CARDOSO2017137}
Jo{ã}o~M.P. Cardoso, Jos{é} Gabriel~F. Coutinho, and Pedro~C. Diniz.
\newblock Chapter 5 - source code transformations and optimizations.
\newblock In Jo{ã}o~M.P. Cardoso, Jos{é} Gabriel~F. Coutinho, and Pedro~C.
  Diniz, editors, {\em Embedded Computing for High Performance}, pages
  137--183. Morgan Kaufmann, Boston, 2017.

\bibitem{dickson2011importance}
Neil~G Dickson, Kamran Karimi, and Firas Hamze.
\newblock Importance of explicit vectorization for cpu and gpu software
  performance.
\newblock {\em Journal of Computational Physics}, 230(13):5383--5398, 2011.

\bibitem{ebner2008generalized}
Dietmar Ebner, Florian Brandner, Bernhard Scholz, Andreas Krall, Peter
  Wiedermann, and Albrecht Kadlec.
\newblock Generalized instruction selection using ssa-graphs.
\newblock In {\em Proceedings of the 2008 ACM SIGPLAN-SIGBED conference on
  Languages, compilers, and tools for embedded systems}, pages 31--40, 2008.

\bibitem{eldridge2021mlir}
Schuyler Eldridge, Prithayan Barua, Aliaksei Chapyzhenka, Adam Izraelevitz,
  Jack Koenig, Chris Lattner, Andrew Lenharth, George Leontiev, Fabian Schuiki,
  Ram Sunder, et~al.
\newblock Mlir as hardware compiler infrastructure.
\newblock In {\em Workshop on Open-Source EDA Technology (WOSET)}, 2021.

\bibitem{frostig2018compiling}
Roy Frostig, Matthew~James Johnson, and Chris Leary.
\newblock Compiling machine learning programs via high-level tracing.
\newblock {\em Systems for Machine Learning}, 4(9), 2018.

\bibitem{hayakawa1948art}
SI~Hayakawa.
\newblock 'the art of plain talk'.
\newblock {\em American Speech}, 23(2):138--141, 1948.

\bibitem{ike2015inside}
Obi Ike-Nwosu.
\newblock Inside the python virtual machine, 2015.

\bibitem{pybind11}
Wenzel Jakob, Jason Rhinelander, and Dean Moldovan.
\newblock pybind11 - seamless operability between c++11 and python, 2016.
\newblock https://github.com/pybind/pybind11.

\bibitem{https://doi.org/10.48550/arxiv.2008.08272}
Tian Jin, Gheorghe-Teodor Bercea, Tung~D. Le, Tong Chen, Gong Su, Haruki Imai,
  Yasushi Negishi, Anh Leu, Kevin O'Brien, Kiyokuni Kawachiya, and Alexandre~E.
  Eichenberger.
\newblock Compiling onnx neural network models using mlir, 2020.

\bibitem{10.14778/3551793.3551801}
Michael Jungmair, Andr\'{e} Kohn, and Jana Giceva.
\newblock Designing an open framework for query optimization and compilation.
\newblock {\em Proc. VLDB Endow.}, 15(11):2389--2401, jul 2022.

\bibitem{nuitka}
{Kay Hayen}.
\newblock Nuitka the python compiler.
\newblock \url{https://nuitka.net/}, 2023.
\newblock [Online; accessed 26-March-2023].

\bibitem{kessenich2018spir}
John Kessenich, Boaz Ouriel, and Raun Krisch.
\newblock Spir-v specification.
\newblock {\em Khronos Group}, 3:17, 2018.

\bibitem{kiselyov2012typed}
Oleg Kiselyov.
\newblock Typed tagless final interpreters.
\newblock {\em Generic and indexed programming: International spring school,
  sSGIP 2010, oxford, uK, march 22-26, 2010, revised lectures}, pages 130--174,
  2012.

\bibitem{lam2015numba}
Siu~Kwan Lam, Antoine Pitrou, and Stanley Seibert.
\newblock Numba: A llvm-based python jit compiler.
\newblock In {\em Proceedings of the Second Workshop on the LLVM Compiler
  Infrastructure in HPC}, pages 1--6, 2015.

\bibitem{10.1145/2833157.2833162}
Siu~Kwan Lam, Antoine Pitrou, and Stanley Seibert.
\newblock Numba: A llvm-based python jit compiler.
\newblock In {\em Proceedings of the Second Workshop on the LLVM Compiler
  Infrastructure in HPC}, LLVM '15, New York, NY, USA, 2015. Association for
  Computing Machinery.

\bibitem{lattner2004llvm}
Chris Lattner and Vikram Adve.
\newblock Llvm: A compilation framework for lifelong program analysis \&
  transformation.
\newblock In {\em International symposium on code generation and optimization,
  2004. CGO 2004.}, pages 75--86. IEEE, 2004.

\bibitem{https://doi.org/10.48550/arxiv.2002.11054}
Chris Lattner, Mehdi Amini, Uday Bondhugula, Albert Cohen, Andy Davis, Jacques
  Pienaar, River Riddle, Tatiana Shpeisman, Nicolas Vasilache, and Oleksandr
  Zinenko.
\newblock Mlir: A compiler infrastructure for the end of moore's law, 2020.

\bibitem{ladderdialect}
{Lei Zhang}.
\newblock Mlir codegen dialects for machine learning compilers.
\newblock
  \url{https://www.lei.chat/posts/mlir-codegen-dialects-for-machine-learning-compilers/},
  2022.
\newblock [Online; accessed 26-March-2023].

\bibitem{li2009note}
Yinan Li, Jack Dongarra, and Stanimire Tomov.
\newblock A note on auto-tuning gemm for gpus.
\newblock In {\em Computational Science--ICCS 2009: 9th International
  Conference Baton Rouge, LA, USA, May 25-27, 2009 Proceedings, Part I 9},
  pages 884--892. Springer, 2009.

\bibitem{9605269}
Alexander McCaskey and Thien Nguyen.
\newblock A mlir dialect for quantum assembly languages.
\newblock In {\em 2021 IEEE International Conference on Quantum Computing and
  Engineering (QCE)}, pages 255--264, 2021.

\bibitem{parr1995antlr}
Terence~J. Parr and Russell~W. Quong.
\newblock Antlr: A predicated-ll (k) parser generator.
\newblock {\em Software: Practice and Experience}, 25(7):789--810, 1995.

\bibitem{paszke2017automatic}
Adam Paszke, Sam Gross, Soumith Chintala, Gregory Chanan, Edward Yang, Zachary
  DeVito, Zeming Lin, Alban Desmaison, Luca Antiga, and Adam Lerer.
\newblock Automatic differentiation in {PyTorch}.
\newblock 2017.

\bibitem{nevergrad}
J.~Rapin and O.~Teytaud.
\newblock {Nevergrad - A gradient-free optimization platform}.
\newblock \url{https://GitHub.com/FacebookResearch/Nevergrad}, 2018.

\bibitem{10.1145/3391902}
Nico Reissmann, Jan~Christian Meyer, Helge Bahmann, and Magnus Sj\"{a}lander.
\newblock Rvsdg: An intermediate representation for optimizing compilers.
\newblock {\em ACM Trans. Embed. Comput. Syst.}, 19(6), dec 2020.

\bibitem{10.1145/3578360.3580275}
Ariya Shajii, Gabriel Ramirez, Haris Smajlovi\'{c}, Jessica Ray, Bonnie Berger,
  Saman Amarasinghe, and Ibrahim Numanagi\'{c}.
\newblock Codon: A compiler for high-performance pythonic applications and
  dsls.
\newblock In {\em Proceedings of the 32nd ACM SIGPLAN International Conference
  on Compiler Construction}, CC 2023, pages 191--202, New York, NY, USA, 2023.
  Association for Computing Machinery.

\bibitem{smith2002introducing}
Graeme Smith.
\newblock Introducing reference semantics via refinement.
\newblock In {\em Formal Methods and Software Engineering: 4th International
  Conference on Formal Engineering Methods, ICFEM 2002 Shanghai, China, October
  21--25, 2002 Proceedings 4}, pages 588--599. Springer, 2002.

\bibitem{troelsen2017philosophy}
Andrew Troelsen, Philip Japikse, Andrew Troelsen, and Philip Japikse.
\newblock The philosophy of. net core.
\newblock {\em Pro C\# 7: With. NET and. NET Core}, pages 1245--1253, 2017.

\bibitem{https://doi.org/10.48550/arxiv.2202.03293}
Nicolas Vasilache, Oleksandr Zinenko, Aart J.~C. Bik, Mahesh Ravishankar,
  Thomas Raoux, Alexander Belyaev, Matthias Springer, Tobias Gysi, Diego
  Caballero, Stephan Herhut, Stella Laurenzo, and Albert Cohen.
\newblock Composable and modular code generation in mlir: A structured and
  retargetable approach to tensor compiler construction, 2022.

\bibitem{10.1145/3178372.3179509}
Jie Zhao, Michael Kruse, and Albert Cohen.
\newblock A polyhedral compilation framework for loops with dynamic
  data-dependent bounds.
\newblock In {\em Proceedings of the 27th International Conference on Compiler
  Construction}, CC 2018, pages 14--24, New York, NY, USA, 2018. Association
  for Computing Machinery.

\bibitem{zimmerman2022langcc}
Joe Zimmerman.
\newblock langcc: A next-generation compiler compiler, 2022.

\end{thebibliography}

\end{document}